\documentclass[aps,prl,twocolumn,showpacs,groupedaddress]{revtex4}
\usepackage{graphicx,epsfig,dcolumn,afterpage}
\usepackage[latin1]{inputenc}

\usepackage{color}
\usepackage{colortbl}

\newcommand{\Wsqcm}{\,\mbox{$\textrm{W} / \textrm{cm}^2$}}
\newcommand{\eVWsqcm}{\,\mbox{$\textrm{eV} / (\textrm{W} / \textrm{cm}^2)$}}
\newcommand{\GVm}{\,\mbox{$\textrm{GV} / \textrm{m}$}}

\newcommand{\phiEff}{\phi_{\mathrm{eff}}}

\newcommand{\Fdc}{F_{\mathrm{dc}}}

\begin{document}

\title{Strong-field above-threshold photoemission from sharp metal tips}

\author{Markus Schenk}
\affiliation{Max-Planck-Institut f{\"u}r Quantenoptik, \\ Hans-Kopfermann-Str. 1, 85748 Garching, Germany}
\author{Michael Kr{\"u}ger}
\affiliation{Max-Planck-Institut f{\"u}r Quantenoptik, \\ Hans-Kopfermann-Str. 1, 85748 Garching, Germany}
\author{Peter Hommelhoff}
\email{peter.hommelhoff@mpq.mpg.de}
\affiliation{Max-Planck-Institut f{\"u}r Quantenoptik, \\ Hans-Kopfermann-Str. 1, 85748 Garching, Germany}

\date{\today}

\begin{abstract}
We present energy-resolved measurements of electron emission from sharp metal tips driven with low energy pulses from a few-cycle laser oscillator. We observe above-threshold photoemission with a photon order of up to 9. At a laser intensity of $\sim$$2 \cdot 10^{11}$\Wsqcm\ suppression of the lowest order peak occurs, indicating the onset of strong-field effects. We also observe peak shifting linearly with intensity with a slope of around $-1.8 \, \textrm{eV} / (10^{12}\,\textrm{W} / \textrm{cm}^2)$. We attribute the magnitude of the laser field effects to field enhancement taking place at the tip's surface.
\end{abstract}


\maketitle

\vspace{1cm}

An intriguing effect in the realm of laser-matter interaction describes the ionization of atoms with more photons than necessary, a phenomenon called above-threshold ionization (ATI).
First observed in 1979~\cite{Agostini1979}, it still attracts considerable interest because of the rich, fundamental, and universal underlying physics  (for reviews see, e.g.,~\cite{Eberly1991, DeloneKrainovMultiphoton1994, MilosevicBecker2006}) and because of its accompanying effect of high-harmonic generation that opened the field of attosecond science (see, e.g.,~\cite{Bucksbaum2007, Krausz2009}). Above-threshold photoemission (ATP) is the solid-state analog of ATI. Whereas a vast amount of literature exists on ATI, ATP has been studied to a much lesser extent and so far only from flat surfaces~\cite{Luan1989,Fann1991,Farkas1993,Aeschlimann1995, Banfi2005, Bisio2006, Saathoff2008}. Here we demonstrate high photon orders and strong-field effects in ATP, both being clearly visible in the peak structure of the electron spectrum.

When femtosecond laser pulses are focused on a sharp metal tip in the presence of a dc electric field, electrons are laser-emitted in the tip-pointing direction. Such systems have been shown to represent ultrafast laser-driven nanoscale electron emitters~\cite{Hommelhoff_PRL2006, Hommelhoff_PRL2006_2, Ropers2007, Barwick_Batelaan_NJP2007}, which have already been used to prove the dispersionless nature of the Aharonov-Bohm effect~\cite{Caprez_Batelaan_PRL2007} and to build a new nanometric imaging tool~\cite{Ropers2007}. Because of their high brightness laser-triggered nanoemitters bear much potential as low-emittance electron sources for future free electron lasers~\cite{Ganter2008,Tsujino2008,Tsujino2009}. Due to the nanometric dimensions of the emitting tip, surface electromagnetic waves (surface plasmon-polaritons) can be excited and have been accurately studied by observing the electron emission pattern associated with the plasmonic surface electric field~\cite{Yanagisawa2009, Yanagisawa2010}.  Such surface excitations can lead to an enhancement of the incident laser  intensity by one to two orders of magnitude, thereby substantially lowering the requirements on the laser system. We will show later that it is most likely this effect that enables us to observe strong-field effects even though we only work with a low power laser oscillator.

In this letter we focus few-cycle femtosecond laser pulses on a sharp tungsten tip and measure the energy of the emitted electrons. Until hitherto energy resolved measurements from sharp tips with pulsed lasers have only been studied in a different parameter range and focusing on different physical aspects~\cite{Tsujino2009, Hilbert2009}.

\begin{figure}
\centerline{\includegraphics[width=8cm]{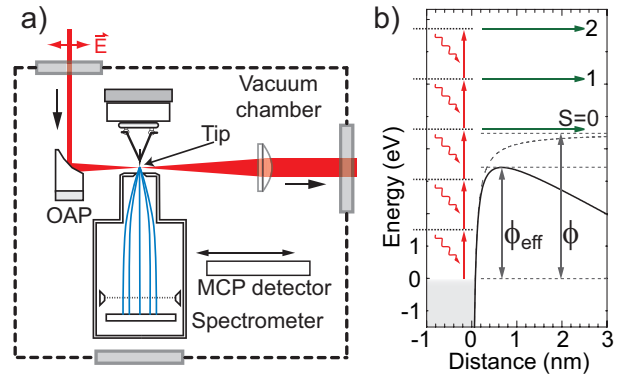}}
\caption{\label{fig:Setup}(Color online). a) Sketch of the experimental setup (not to scale). The retarding field spectrometer can be moved away from the tip (out of plane) to make room for the imaging MCP detector. The laser output viewport is used for alignment. OAP: off-axis parabolic focusing mirror. b) Energy diagram of the metal-vacuum interface with respect to the Fermi energy with an applied dc field of $0.8\GVm$.}
\end{figure}

We generate $6.5\,$fs laser pulses in a Kerr-lens mode-locked Ti:sapphire laser with 80\,MHz repetition rate at a center wavelength of $\sim$800\,nm. The pulses are focused onto a tungsten tip with a gold coated $90^{\circ}$ off-axis parabolic mirror with an effective focal length of 15\,mm (Fig.~\ref{fig:Setup}). We obtain a spot size of $(2.4 \pm 0.2)\,\mu$m ($1/e^2$ intensity radius). Tip and focusing mirror are mounted inside a vacuum chamber with a base pressure of $3 \cdot 10^{-8}\,$Pa. The tip is mounted on a piezo-controlled 3d translation stage so that it can be moved into the region of highest light intensity.  The electric polarization vector of the light field is parallel to the tip pointing direction.

We detect emitted electrons and measure their energy with a retarding field spectrometer with an overall detection efficiency of about $0.1 \dots 1\,\%$ of the emitted electrons. Alternatively, an imaging two stage micro-channel plate (MCP) detector can be moved in front of the tip at a distance of about 4\,cm to observe the spatial emission pattern.  Spectrometer as well as MCP allow detection of single electrons (pulse counting mode) as well as detection of dc currents for larger currents.  In the energy-resolved measurement mode the tip is grounded and the entrance aperture of the spectrometer (around $2\,$mm away from the tip) lies at a variable high positive voltage, whereas in the spatially resolved measurement mode the tip is at high negative voltage and the MCP's front side at ground potential. The spectrometer's spectral resolution is $\sim$80\,meV inferred from a dc field emission curve and constant over a large dc voltage range. All data shown in this letter are recorded with on average less than one electron per laser pulse emitted from the tip in order to avoid Coulomb repulsion effects.

The tungsten tip is electrochemically etched from W(310) single crystal wire. The (310) orientation of tungsten has the lowest work function of the prominent low-index planes~\cite{Kawano2008}. Therefore, the majority of electrons is emitted in the forward direction into a single cone with $\le 20^{\circ}$ opening angle (FWHM), which is why we chose W(310). Before every measurement we field-evaporate the tip at a high positive voltage to obtain a clean surface~\cite{Tsong1990}.

\begin{figure}
\centerline{\includegraphics[width=8.8cm]{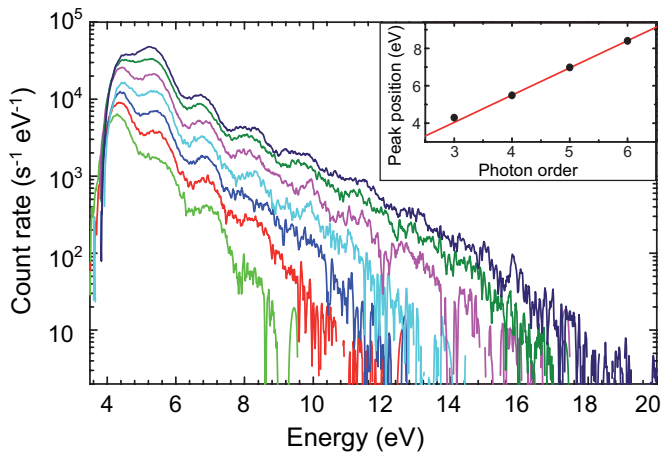}}
\caption{\label{fig:ATPplot}(Color online). Electron count rate as a function of the electron energy. From bottom to top the curves are taken at laser intensities of $\{0.9, 1.2, 1.4, 1.6, 1.9, 2.1, 2.3\} \cdot 10^{11}$\Wsqcm.  Inset: The positions of the peaks of the curve at $1.2 \cdot 10^{11}$\Wsqcm. Details in the text.}
\end{figure}

In Fig.~\ref{fig:ATPplot} we show electron spectra obtained at a (cycle-averaged) peak intensity of $0.9 \dots 2.3 \cdot 10^{11}$\Wsqcm\ (average power of $9 \dots 22\,$mW) and a dc field of $\Fdc = 0.8\,$\GVm\ at the tip apex. The plot is obtained by measuring the electron count rate as a function of retardation voltage and subsequent differentiation and smoothing. The energy axis's origin, here defined to be the Fermi energy, is determined with the help of the dc field emission peak in a separate measurement. Clearly, peaks on top of an overall exponential decay are visible, together with a cut-on. In the following we will discuss the origin of these features.

Due to the Schottky effect~\cite{FurseyBook2005} the effective barrier height is reduced to $\phiEff = \phi - \sqrt{e^3  \Fdc / (4 \pi \epsilon_0)}$, where $\phi = 4.35\,$eV is the work function of W(310), $\epsilon_0$ is the vacuum permittivity, and $e$ the electron charge [Fig.~\ref{fig:Setup}~b)]. For sharp tips, $\Fdc = U / (k r),$ where $U$ is the voltage between tip and anode and $r$ the tip radius. $k$ is a factor that takes into account the shielding of the field due to the presence of the tip shank and approximately equals 5~\cite{FurseyBook2005}. For $r \approx 38\,$nm as used here a moderate voltage of 150\,V suffices to generate a field of $0.8$\GVm\ at the tip apex, thus $\phiEff = 3.3$\,eV.

In the experiment we observe that $\phiEff$ grows larger within about one hour after field evaporation. The effectively increased $\phiEff$ is directly visible in the cut-on in Fig.~\ref{fig:ATPplot}. From the figure we estimate $\phiEff \approx 4$\,eV. Right after field evaporation, $\phiEff$ is as expected. Even though it is desirable to understand the exact mechanism that leads to the fast change in $\phiEff$, we do not observe any other changes in the spectra than a shift of the cut-on.

For emission over the barrier the threshold number of photons necessary is $K = \langle \phiEff / (\hbar \omega) +1 \rangle$ where $\langle \ \rangle$ denotes the integer part and $\hbar  \omega \approx 1.56\,$eV. Thus, here three photons are required to lift an electron over the barrier [Fig.~\ref{fig:Setup}~b)]. In the experiment (Fig.~\ref{fig:ATPplot}), we observe the dominant lowest-order peak at around $4.3\,$eV in the five low intensity curves. In addition to this peak, which represents the well-known multi-photon emission, several more peaks are visible with energies $(K+S) \hbar \omega.$ Here $S$ is the above-threshold order and reaches up to 6, i.e. processes with photon order up to $(K+S) = 9$ are visible.

The inset of Fig.~\ref{fig:ATPplot} shows the energy of the peaks visible in Fig.~\ref{fig:ATPplot} at an intensity of $1.2 \cdot 10^{11}$\Wsqcm. A linear fit reveals that the peaks are equally spaced by $\sim$$1.46$\,eV, clearly indicating the ATP nature of the process. The three-photon peak position is located slightly above the fit curve, which includes the higher photon orders only. This deviation is due to the barrier which partially cuts this peak. If we extrapolate the linear fit to zero photon order we see that it intersects the energy axis at $\sim 0.3\,$eV below the Fermi level. At $-0.4\,$eV the local density of states of tungsten exhibits a peak for W(310)~\cite{Plummer1970}. We conclude that the majority of emitted electrons originate from there and in general from the vicinity of the Fermi level.

A qualitative change in the spectra occurs if the laser intensity is increased from $1.9$ to $2.3 \cdot 10^{11}\Wsqcm$: The yield of the $S=1$ peak exceeds that of the $S=0$ peak. In ATI, this effect is well-known and has been called {\it peak suppression} (also known as {\it threshold shifting} or onset of {\it channel closing})~\cite{Eberly1991,DeloneKrainovMultiphoton1994}. It mainly reflects the fact that the continuum experiences an AC Stark shift due to the presence of the laser field so that the ionization potential is effectively increased~\cite{Muller1983,Muller1988,Mulser1993}. If the light shifted potential barrier exceeds the energy of a given photon order, the corresponding peak will be suppressed. We conclude that, in close analogy, here the $S=0$ peak is becoming suppressed with increasing laser intensity due to the increasingly light shifted continuum.

Following~\cite{Saathoff2008}, we argue that the initial state's light shift is negligible against the continuum light shift. Tungsten is a d-band transition metal implying that electrons near the Fermi level are predominantly found in localized d-bands. The localization of the initial state electrons leads to a  behavior similar to ATI from atoms where the ground state light shift is usually much smaller than that of the continuum and therefore negligible.

With this assumption the threshold number of photons is given by $K =  \langle [ \phiEff  + U_p]/(\hbar \omega) +1  \rangle$ with the ponderomotive energy $U_p=e^2 I_L / (2 c \epsilon_0 m \omega^2),$ where $I_L$ and $\omega$ are the peak intensity and carrier (circular) frequency of the driving laser pulse, $c$ the vacuum speed of light, and $m$ the electron mass. $U_p$ represents the quiver energy of a free electron in a laser field. This means classically that the to-be-liberated electron has not only to provide the energy to overcome the barrier but also the quiver energy in the laser field. Speaking in the dressed-states picture, the continuum levels are AC Stark shifted by $U_p$~\cite{DeloneKrainovMultiphoton1994,Mulser1993}.

From data sets such as the one shown in Fig.~\ref{fig:ATPplot} we infer that the critical intensity $I_c$ at which the $S=1$ peak reaches the $S=0$ peak in amplitude is in the range of $0.5 \dots 2.0  \cdot 10^{11} \Wsqcm$ depending on $\Fdc$ via $\phiEff$. This means that here, surprisingly, as little as  10\,mW  of average power of an 80\,MHz few-cycle oscillator suffice to reach the regime in which strong-field effects start to dominate.

\begin{figure}
\centerline{\includegraphics[width=5cm]{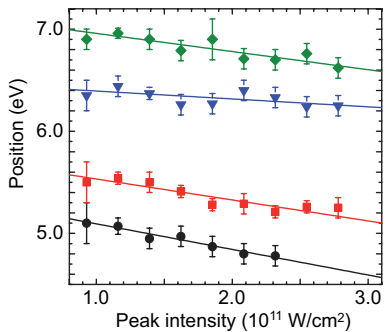}}
\caption{\label{fig:PeakShift}(Color online). Positions of the $(K+S) = 4$ and 5 maxima (squares and diamonds) and of the next lower minima (balls and triangles) as extracted from Fig.~\ref{fig:ATPplot} as a function of laser intensity. The slopes are in the range of $-2.5 \dots -0.8 \cdot 10^{-12}\,\eVWsqcm.$}
\end{figure}

Closely related to the peak suppression effect is a shift of the features in Fig.~\ref{fig:ATPplot} with laser intensity. We plot the positions of maxima and minima of Fig.~\ref{fig:ATPplot} as a function of laser intensity in Fig.~\ref{fig:PeakShift}. Maxima and minima downshift in energy roughly linearly with increasing laser intensity presumably also owing to the light shift. When we fit straight lines to these data we obtain a mean slope of $- 1.8 \cdot 10^{-12}$\eVWsqcm.  This slope should be compared to the theoretically expected one, which is  $- U_p / I_L = - 5.5 \cdot 10^{-14}$\eVWsqcm, again assuming that the light shift of the continuum equals $U_p$ and that the initial state is unaffected by the light. The experimentally observed slope is much larger than the theoretically expected one. This can result from a continuum light shift larger than expected and/or from a light shift of the initial state (with opposite sign) and/or from an enhancement of the laser intensity in close proximity of the tip. If we assume that the difference is due only to field enhancement, we obtain a field enhancement factor of $5.7$ by taking the square root of the ratio of the slopes. This factor is slightly larger but comparable to what can be expected and has been observed previously for systems similar to this one~\cite{Martin2001, Hommelhoff_PRL2006, Hommelhoff_PRL2006_2}.

Note that because of the shortness of the laser pulse the emitted electron does not ``surf'' down a ponderomotive potential~\cite{Bucksbaum1987}, which would render the peak shift unobservable, but returns the ponderomotive energy to the radiation field. This is in fact why the peak shift is observable, and why the notion of the AC-Stark shift seems appropriate~\cite{Eberly1991, Mulser1993}.

We take this agreement as evidence for the following qualitative picture: The continuum is light shifted by an amount equal to the ponderomotive energy, a ground state shift is negligible, and field enhancement at the tip takes place. However, from Fig.~\ref{fig:PeakShift} it is apparent that neither maxima nor minima downshift perfectly linear with laser intensity. Also, the slopes of the different maxima and minima differ by about a factor of three. The reasons for both of these points are thus far not clear to us and need further investigation. Furthermore, an evaluation of a possible peak shift of the higher order peaks similar to what is shown in  Fig.~\ref{fig:PeakShift} is not possible because of poor statistics.

From ATI it is well known that the peak structure disappears when the atoms are driven with few-cycle laser pulses~\cite{Grasbon2003}. This is attributed to the fact that in the gas jet atoms are exposed to regions of different peak intensities and experience a different light shift, a phenomenon called {\it volume effect}. In contrast, here the ATP peak structure is clearly visible even though the laser pulse duration is well in the few-cycle regime. This is because the tip's electron source area is much smaller than the laser spot size so that the peak intensity is well-defined. Note that the source area can be shrunk down to the size of a single atom~\cite{Fink88, Hommelhoff_Ultramic2009} so that precision studies of ATI and ATP with single atoms or single attached molecules should be feasible. A small electron emission probability will be mitigated by the high repetition rate of the driving laser oscillator.

It will be interesting to investigate the time scales involved in the processes. Here we only note that the observation of ATP together with strong-field effects can be taken as evidence for an electron emission duration of around or less than the laser pulse duration. The finite lifetime of the plasmonic surface excitation could extend the window during which external field effects can take place, but in tungsten the surface excitation should be damped out within less than 10\,fs due to the comparatively high imaginary part of the dielectric constant~{\cite{Stietz2000,Yanagisawa2010}.

Including the field enhancement factor, the Keldysh parameter $\gamma$~\cite{DeloneKrainovMultiphoton1994} lies in the range of $2.2  \dots 3.5$ for the experimental parameters occurring in this letter, indicating that the emission process falls in the transition regime of tunneling and multiphoton picture, leaning towards the latter. However, note that the presence of an additional static electric field is not included in the Keldysh theory, rendering the notion of $\gamma$ doubtful here. In numerical simulations we see that already for a Keldysh parameter as large as $5$ a sizeable component of the electron current resolves the temporal structure of the optical electrical field.

To summarize, we have observed above-threshold photoemission from sharp tungsten tips with a photon order of up to 9. In addition, we have observed peak suppression and peak shifting. To the best of our knowledge these strong-field field effects have not been reported before in ATP. We were able to achieve these results with a low power, high repetition rate oscillator, most likely owing to field enhancement due to excitation of surface electromagnetic waves on the highly curved tip surface.

As an outlook we mention that we have clear evidence for the recollision of the liberated electron with the parent tip, an effect essential for the generation of high-harmonic generation~\cite{Bucksbaum2007,Krausz2009}. Analogous to what is known from ATI~\cite{Becker2002} we have observed a pronounced intensity-dependent kink between two exponential slopes in the yield-vs.-energy curve (to be published). Therefore, the combination of a nanoemitter with few-cycle laser pulses will also allow studying other strong-field effects such as recollision physics and high-harmonic generation, and this without focal averaging, which usually makes the analysis of ATI data taken in a gas jet cumbersome. In addition, a dc field comparable in strength to the peak laser field can be applied to the tip, which allows tuning of both the effective barrier height and the dynamics of the liberated electron, thereby offering a new parameter to steer the electronic motion.

We thank Matthias Kling for insightful discussions, the H{\"a}nsch group for fruitful interaction, Karl Linner and Wolfgang Simon for excellent technical support during the setup phase and the Max-Planck Society and European Union (FP7) for funding.


\end{document}